\documentclass[10pt,conference]{IEEEtran}

\usepackage{cite}
\usepackage{amsmath,amssymb,amsfonts}
\usepackage{algorithmic}
\usepackage{graphicx}
\PassOptionsToPackage{table,dvipsnames,svgnames}{xcolor}
\usepackage{xcolor}
\usepackage[final]{microtype}
\usepackage[italic]{mathastext}
\usepackage{libertine}
\usepackage[T1]{fontenc}
\usepackage{textcomp}
\usepackage[varqu,varl]{zi4}
\usepackage[all]{nowidow}
\usepackage[keeplastbox]{flushend}
\usepackage{fancyhdr}

\usepackage{expl3}
\usepackage{graphicx}
\usepackage{tikz}
\usetikzlibrary{positioning, shapes.multipart, arrows.meta, calc}
\usepackage{amsmath}
\usepackage{amssymb}
\usepackage{booktabs}  
\usepackage{paralist}
\newcommand{\para}[1]{\smallskip\noindent\textbf{{#1.}}}
\usepackage[utf8]{inputenc}    
\usepackage{enumitem}    
\usepackage{comment}
\usepackage{subcaption}
\PassOptionsToPackage{hyphens}{url}
\usepackage{hyperref}      
\usepackage{cleveref}
\usepackage{tabularx}      
\usepackage{makecell}      
\usepackage{textcomp}      
\usepackage{ragged2e}      
\usepackage{lipsum}
\usepackage{threeparttable}
\crefname{algocf}{algorithm}{algorithms}
\Crefname{algocf}{Algorithm}{Algorithms}
\usepackage[T1]{fontenc} 

\usepackage{caption} 
\usepackage{cprotect}
\usepackage[inkscapelatex=false]{svg} 

\newcolumntype{L}{>{\RaggedRight\arraybackslash}X}
\newcolumntype{C}{>{\centering\arraybackslash}X}

\definecolor{tableheadspec}{HTML}{446688}

\title{Catscan: Visualizing Pipelines of CPU Performance Simulation}

\author{%
  \IEEEauthorblockN{Aaron Lindsay, Nicholas Kelly, Scott Witscher, Mahesh Madhav}
  \IEEEauthorblockA{\emph{Ampere Computing}, Raleigh, NC and Portland, OR}
}

\makeatletter
\renewcommand{\verbatim@font}{\ttfamily\footnotesize\bfseries} 
\makeatother

\hypersetup{
    colorlinks=true,
    linkcolor=ForestGreen,
    citecolor=ForestGreen,
    urlcolor=RoyalBlue,
    filecolor=IndianRed,
    pdftitle={Catscan: Visualizing Pipelines of CPU Performance Simulation}, 
    pdfauthor={Aaron Lindsay, Nicholas Kelly, Scott Witscher, Mahesh Madhav},
    pdfsubject={Visualization}, 
    pdfkeywords={CPU Performance, Microarchitecture, Tools, Visualization, Analysis}, 
    pdfnewwindow=true,
    pdfdisplaydoctitle=true,
    bookmarksopen=true
}

\begin{document}
\maketitle
\pagestyle{plain}

\newcommand{\Red}[1]{{\color{red} #1}}
\newcommand{\ignore}[1]{}
\newcommand{\asm}[1]{\texttt{#1}}
\newcommand{\sys}[1]{\texttt{#1}}
\newcommand{\kw}[1]{\textit{#1}}
\newcommand{\kwb}[1]{\textbf{#1}}
\newcommand{\type}[1]{\textit{#1}}
\newcommand{\Response}[1]{{\color{blue} #1}}
\newcommand{\XXX}[1]{\Red{\textbf{XXX[}#1\textbf{]}}}
\newcommand{\tocite}[1]{\Red{CITE:\cite{#1}}}
\newcommand{\toref}[1]{\Red{REF:\ref{#1}}}
\newcommand{\parasub}[1]{\smallskip\noindent\textit{{#1:}\xspace}}
\newcommand{\mahesh}[1]{\textcolor{purple}{#1}}
\newcommand{\anyone}[1]{\textcolor{blue}{#1}}
\newcommand{\niparagraph}[1]{\noindent\textbf{\textsf{#1}\hspace{0.5em}}}
\newcommand\TODO[1]{\textcolor{red}{TODO: #1}}

\newenvironment{CompactItemize}%
  {\begin{list}{$\blacktriangleright$}%
    {\leftmargin=\parindent \itemsep=2pt \topsep=2pt
     \parsep=0pt \partopsep=0pt}}%
  {\end{list}}
\renewcommand{\labelitemi}{$\blacktriangleright$}

\newcommand{\malloc}{{\texttt{Malloc}}}
\newcommand{\linklist}{{\textsf{Linked-List}}}

\definecolor{miss_event}{RGB}{255, 80, 120}
\definecolor{alloc_event}{RGB}{255, 119, 39}
\definecolor{hit_event}{RGB}{0, 160, 255}
\definecolor{dealloc_event}{RGB}{120, 120, 255}

\DeclareRobustCommand{\misshighlightnarrow}[1]{%
  {%
    \setlength{\fboxsep}{0pt}
    \colorbox{miss_event}{%
      \strut\textcolor{white}{{\fontseries{c}\selectfont\ttfamily #1}}%
    }%
  }%
}

\DeclareRobustCommand{\allochighlightnarrow}[1]{%
  {%
    \setlength{\fboxsep}{0pt}%
    \colorbox{alloc_event}{%
      \strut\textcolor{white}{{\fontseries{c}\selectfont\ttfamily #1}}%
    }%
  }%
}

\DeclareRobustCommand{\hithighlightnarrow}[1]{%
  {%
    \setlength{\fboxsep}{0pt}%
    \colorbox{hit_event}{%
      \strut\textcolor{white}{{\fontseries{c}\selectfont\ttfamily #1}}%
    }%
  }%
}

\DeclareRobustCommand{\deallochighlightnarrow}[1]{%
  {%
    \setlength{\fboxsep}{0pt}%
    \colorbox{dealloc_event}{%
      \strut\textcolor{white}{{\fontseries{c}\selectfont\ttfamily #1}}%
    }%
  }%
}
\begin{abstract}

Processor pipeline visualization tools are routine inside industry CPU teams, but few of them are described or released publicly. As a result, students, researchers, and other practitioners rarely see the tooling that processor architects use to debug performance before silicon. This paper describes two pieces of Ampere Computing's performance-analysis infrastructure that we have released to the community as open source: event streams, a simulator-output format, and Catscan, an interactive viewer built around that format. Event streams record microarchitectural activity as typed events connected by transaction relationships, so a user can move between a symptom and the instruction, uop, or memory transaction that explains it. Catscan uses that structure to support resource- and transaction-oriented views, persistent highlighting, domain-specific search, comparative trace synchronization, and other workflows used during product development. In this paper we report the design choices that survived production use, the limitations we encountered, and the lessons we think are useful for future microarchitectural visualization tools.

\end{abstract}

\section{Introduction}

The relentless increase in CPU complexity, driven by features like out-of-order execution and deep speculation, has made pre-silicon performance analysis and bug diagnosis a formidable challenge. Microarchitectural simulators are indispensable tools in the domain of design space exploration, capable of executing complex workloads and producing a ``dazzling variety of statistics'' \cite{alpha_sim}. These statistics, such as Instructions Per Cycle (IPC) or cache miss rates, are effective at identifying that a performance problem exists. However, they often lack the diagnostic fidelity required to pinpoint \textit{why} it exists, leaving architects to manually sift through text-based logs or spreadsheets of performance counter data.

To bridge this gap between problem detection and root-cause analysis, the field has long turned to execution trace visualization. Early academic and industrial efforts produced tools that could render ``waterfall'' or pipeline diagrams, offering a graphical view of an instruction's journey through the machine's stages over time \cite{alpha_sim, illinois_tracevis, jviewer}, or the contents of buffers as instructions move through \cite{satsim}. This concept evolved into more generalized tools like the Graphical Pipeline Viewer (GPV), which introduced the idea of a generic, text-based ``pipetrace stream'' to remain portable across different simulator backends \cite{michigan_pipeviz}. This early work established the core value proposition that a visual representation of the pipeline is more intuitive for analysis than raw numbers or text logs.

Visualization techniques were significantly advanced by the work of Stolte et al. with their visualization system for superscalar processors \cite{stanford_rivet, stanford_rivet2} which was built on a philosophy of providing an ``overview-plus-detail'' display. This allowed a user to start with a high-level timeline view of pipeline performance, identify regions of poor instruction throughput, and then ``drill down'' to a focused region for in-depth analysis. Their visualization took the form of an animated, cycle-by-cycle reenactment of the pipeline, showing precisely how instructions were reordered and executed, and how functional units were being utilized. This approach proved highly effective, enabling the discovery of subtle timing bugs in simulators that were completely obscured by aggregate statistics. Their body of work solidified the principles of modern CPU pipeline visualization: interactivity, multi-scale views, and the ability to connect low-level events back to high-level performance characteristics \cite{stanford_bosch}. 

However, a common thread unites this foundational work. From the simple stream of GPV \cite{michigan_pipeviz} to the sophisticated overview-plus-detail displays of Rivet \cite{stanford_rivet2}, the input data is fundamentally treated as a time-ordered, flat stream of events. While powerful, this representation loses the inherent causal and hierarchical relationships that define microarchitectural execution. A pipeline stall is not merely an event that occurs at a point in time; it is the \textit{consequence} of a preceding event, such as a cache miss, which itself was \textit{caused} by a specific micro-operation belonging to a particular instruction. However, a flat stream forces the user to mentally reconstruct this critical context during a debug session.

\begin{figure}[!t]
    \centering
    \includegraphics[width=\linewidth]{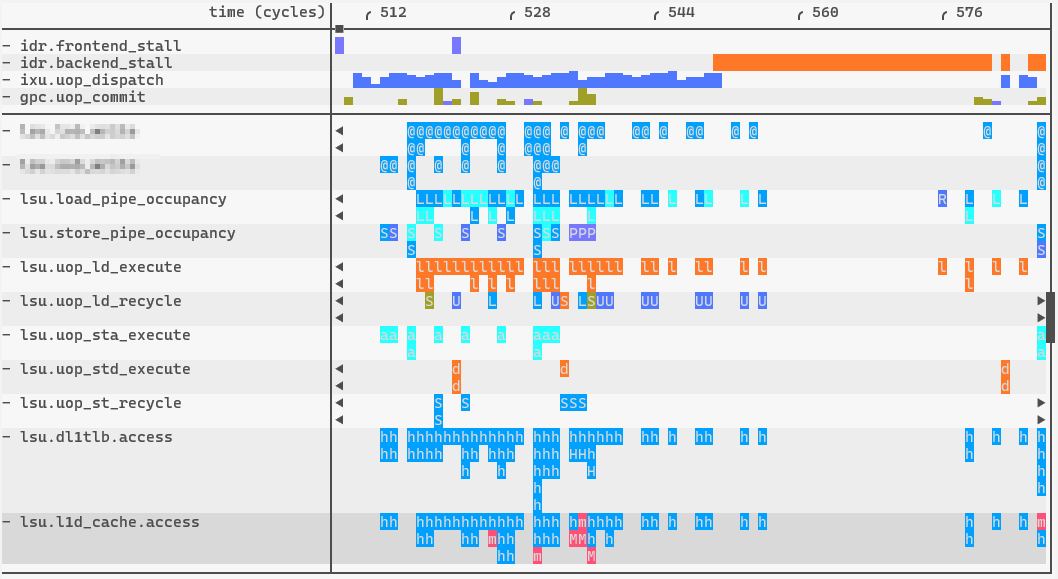}
    \caption{Catscan debugging session demonstrating ``row pinning'' of stall frequencies and other critical event bandwidths (represented by multilevel bars at the top) along with other detailed events in a resource-oriented view. In this case, the backend pipeline is stalled, waiting on a series of L1 cache misses to return data (see the demand misses |\misshighlightnarrow{m}|, and prefetch misses |\misshighlightnarrow{M}|.) }
    \label{fig:catscan_pinning}
\end{figure}

We argue that the structure of the trace data itself is a first-class design problem that has been underexplored. This paper shares our philosophy for representing microarchitectural execution not as a stream of standalone events, but as a \textbf{hierarchical causal graph}. In our model, execution is captured as a directed acyclic graph of transactions. For example, a ``fetch'' transaction spawns a set of ``instruction'' transactions. Each ``instruction'' may in turn spawn multiple ``micro-op'' (uop) transactions. A ``uop,'' upon execution, may then spawn dependent transactions, such as a ``TLB lookup'' or a ``memory access.'' This graph-based representation preserves the crucial parent-child and sibling relationships between operations. The individual events, such as \verb|buffer_allocate|, \verb|cache_lookup|, or \verb|data_bypass|, belong to one of these transactions. The resulting data format is a rich, self-contained description of \textit{what} happened as well as the structural and causal \textit{how} and \textit{why}.

Building on this data representation, we present \textit{Catscan}, our \textbf{interactive visualization tool} designed specifically to leverage this hierarchical structure. Unlike historical tools that primarily allow for filtering by time or event type, Catscan allows an architect to explore more efficiently by using the causal chain. For example, a user can select a high-level transaction, such as a long-latency load instruction, and instantly highlight its entire subgraph of dependent uops and memory operations. This provides an intuitive and powerful debugging workflow that mirrors an architect's mental model of the machine. An example screenshot is shown in \autoref{fig:catscan_pinning}.

What further sets this work apart is its foundation in industrial practice. While the prior art represents invaluable academic exploration, those are not the tools used in industry, and their source code is largely unavailable. In contrast, Catscan and event streams have been used in the exploration, design, and performance verification of multiple generations of Ampere's CPU products \cite{ampere_pv, ampere_mte, ampere_panem}. This paper reports the parts of that infrastructure that proved useful enough to release as open source. We have open-sourced both the hierarchical trace format specification and the Catscan visualization tool to give academic and industrial users a reusable framework for microarchitectural performance analysis, extensible for both processor design and computer architecture education.

This paper does not claim that pipeline visualization is new. Instead, it describes the parts of our infrastructure that proved useful enough to keep, extend, and release as open source. We make three main contributions:
\begin{itemize}
    \item We describe event streams, a compact and extensible format for recording typed microarchitectural events and transaction relationships.
    \item We describe Catscan, a terminal-based visualization tool that uses event streams for resource views, transaction views, persistent highlighting, domain-specific search, and comparative trace synchronization.
    \item We report design lessons from using these tools in an industrial CPU development flow, including where the approach works well and where it remains limited.
\end{itemize}

\para{Artifact availability} The open-source artifacts supporting this paper are available in two public repositories. The Catscan repository contains the interactive event-stream viewer \cite{catscan_repo}. The Performance Streams repository contains the event-stream schema/specification, Python and C++ read/write libraries, supporting command-line tools, examples, and documentation for processing event streams \cite{perf_streams_repo}.

\section{Event Streams}
\label{sec:event_streams}

Inside Ampere, event streams are the common format for performance development and debug. Microarchitectural events emitted by software performance models, RTL simulations, and emulation all use this same format. We have built a significant collection of tools to create, filter, process, summarize, compare, and visualize event streams. Using one representation across sources matters in practice: during performance verification, architects can compare a performance model and an RTL run with the same analysis stack instead of learning different tools for each domain.

\subsection{Format}

The event stream format is intentionally lightweight. Its simplicity is what makes it useful across different simulators and levels of modeling detail. The base format records definitions, events, values, and relationships, while each producer defines the domain-specific event names and value meanings it needs.

The format does not prescribe a fixed ISA, microarchitecture, pipeline taxonomy, or simulation style. A stream begins with self-contained definitions, enumeration mappings, and parameters that give stable names and meanings to the events and values used later in the stream (see \autoref{table:definition}). The producer then writes timestamped events, such as \verb|decode| or \verb|commit|, as simulation activity occurs; each event refers to one of those definitions, carries zero or more values (such as the program counter or uop class), and may include relationships used to reconstruct transactions (\autoref{table:event}; \S\ref{sec:transactions}). The producer decides which architectural or microarchitectural behaviors are worth exposing through these definitions, events, values, and relationships.

\begin{table}
\centering
\begin{tabular}{|l|l|}
    \hline
    \textbf{Field} & \textbf{Description} \\
    \hline
    \verb|kind| & Kind of definition (event or value) \\
    \verb|id| & ID of definition \\
    \verb|name| & Name of definition \\
    \verb|description| & Description for definition \\
    \verb|enumeration_id| & Enumeration ID (if is enumeration) \\
    \hline
\end{tabular}
\caption{Definition schema}
\label{table:definition}
\end{table}

\begin{table}
\centering
\begin{tabular}{|l|l|}
    \hline
    \textbf{Field} & \textbf{Description} \\
    \hline
    \verb|definition_id| & Event definition \\
    \verb|id| & Unique ID of event \\
    \verb|time| & Time event occurred (normally picoseconds) \\
    \verb|values| & Map of data values associated with event \\
    \hline
\end{tabular}
\caption{Event schema}
\label{table:event}
\end{table}

We use Protocol Buffers \cite{protobuf} when a binary representation is needed, e.g. saving event streams to disk or streaming them between processes. Protocol Buffers are simple to use, reasonably efficient in space (variable-width fields based upon values) and time, have mature bindings for common languages, and compress well with external tools. Record sizes depend on field widths, values, and producer choices, but the common event records in our streams are compact: an event without attached values is typically on the order of 6-16 bytes before external compression. Inspecting a representative, uncompressed stream yields an average event record size of 27 bytes, which is reduced to only 2 bytes when compressed.

Because event streams do not prescribe a particular ISA, microarchitecture, or simulator structure, we expect them to fit a broad range of existing and future simulators without requiring complex changes. The cleanest integration path is to emit event streams directly from the simulator using the released C++ or Python libraries. Another path is a standalone converter from an existing simulator's trace, assuming the existing trace format has enough timing and identity information to reconstruct useful events and transactions. In both cases, the simulator-specific work is primarily choosing a useful event vocabulary, naming convention, and set of transaction relationships rather than the complex tooling which would be required if using a less flexible event format.

\subsection{Conventions}

\subsubsection{Transactions} \label{sec:transactions}

A core extension of the event stream format is the concept of a transaction. Transactions form a directed acyclic graph (DAG) that captures causal relationships between microarchitectural operations. An instruction transaction, for example, can be the parent of multiple uop transactions, which in turn may spawn descendant transactions for memory accesses or other sub-tasks. This hierarchy allows human analysts and automated tools to traverse a causal chain directly. A question like ``Which uop was responsible for this instruction's long commit latency, and why?'' becomes a graph query rather than a manual reconstruction from unrelated timestamps.

This graph is constructed through conventions layered on top of the base event format. A new transaction begins with a \verb|start_transaction| event and ends with an \verb|end_transaction| event. Each event belonging to a transaction is tagged with a unique \verb|txid| value. A parent-child relationship is established when a \verb|start_transaction| event includes a \verb|parent| value referencing the \verb|txid| of its immediate ancestor. These primitives let a producer write a linear stream while allowing consumers to reconstruct the transaction graph.

A key design choice is that transaction types are usually inferred from their constituent events rather than explicitly declared with a type field. In programming-language terms, transactions are ``duck-typed'': a transaction is identified by what it does, not by a fixed label. This flexibility is useful in a production environment where event streams come from sources with different levels of microarchitectural detail. A high-fidelity model might represent a memory request as a distinct child transaction of a load uop to model memory-controller contention. A lower-fidelity model might instead attach the memory access events directly to the load uop transaction. Duck-typing lets tools handle both representations and still support model-tolerant queries, such as finding the program counter of the instruction whose descendant transactions contain the most L3 cache misses.

\subsubsection{Hierarchy}

Another useful convention is hierarchical naming for events and values. We group events by module and separate hierarchy levels with periods. For example, a TAGE array lookup might be named \verb|core_0.bpu.tage.array_3.lookup|. This convention makes related events easy to specify with glob-like matches in configuration files or on the command line, and it matches how architects already tend to think about modules within a design.

\subsection{Tooling}

Recording events is only useful if the resulting data can be analyzed in realistic debug workflows. The simplicity and flexibility of the event stream format and transaction conventions encouraged the creation of a family of tools: filters by time, event count, event name, and glob-like patterns; text printers for humans and scripts; summary and metric-reporting tools; comparison tools; and Catscan itself. The smaller tools compose well, much like UNIX command-line utilities, which has made it easy for architects to build temporary workflows during debug. We also provide Python and C++ libraries for reading and writing event streams, along with API documentation and examples, so new producers and analysis tools do not need to implement the wire format from scratch \cite{perf_streams_repo}.






\section{Early Visualization}
\label{sec:early_visualization}

When Ampere Computing's performance architecture team was formed in 2018, we did not inherit a performance model or visualization stack. That provided a rare opportunity to build our infrastructure using past lessons learned, but we also had some practical constraints --- we were a small team with limited resources and product deadlines, so each tool had to quickly justify its cost and help architects debug performance.

This context shaped our visualization tooling. We started with simple views that could be built quickly on top of event streams, kept what proved useful, and replaced tools when their limitations became clear. This section describes this progression because it explains several Catscan design choices, including the need to support both resource- and transaction-oriented debugging and to use an intermediate representation from which both views can be derived.

\subsection{Text Diagrams}

Pipeline visualization first proved its value for us through simple text output. Our earliest viewer converted an event stream into a static waterfall diagram meant to be opened in a text editor such as Vim. Each row represented an instruction transaction, time advanced from left to right, and each CPU cycle occupied three characters so multiple events in the same cycle could be shown. We refer to this style as a ``waterfall'' or ``transactional'' view: events are grouped by transaction on the Y-axis and by time on the X-axis.

\begin{figure}[!th]
    \includegraphics[width=\columnwidth]{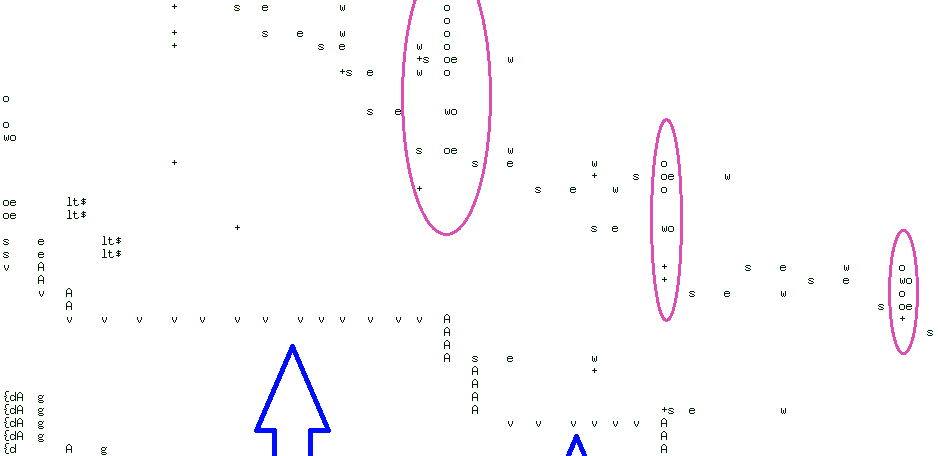}
    \cprotect\caption{An early, static text-based visualization of the pipeline rendered in a text editor. The view has time on the x-axis and individual uops on the y-axis. It captures the diagnosis of a uop scheduler bottleneck, where coarse-grained group deallocations (\verb|o|, circled) lead to frequent allocation stalls (\verb|v|, indicated by blue arrows), which are then followed by new allocations (\verb|A|).}
    \label{fig:stall_2019}
\end{figure}

A compelling example demonstrated the value of visualization when the core microarchitecture was still early in its definition. The team was tasked with identifying performance bottlenecks in the early design of the AmpereOne® core. The initial design implemented a grouping policy where scheduler entries were deallocated in blocks of four, and a block was only freed when all four of its uops had completed execution.

Visualization quickly pointed out a flaw in the geometry of the uop scheduler. The text-based visualization, shown in \autoref{fig:stall_2019}, immediately revealed a recurring pattern of allocation stalls near the front of the pipeline, represented by \verb|v| characters. These stalls persisted until a full group of four older uops completed, at which point a group deallocation (\verb|o|) would free up resources, allowing a burst of new allocations (\verb|A|). The diagram showed that coarse-grained deallocation was wasting scheduler capacity and creating avoidable front-end stalls. This clear visual evidence directly led to changing the microarchitecture to use finer-grained entry management, which improved scheduler utilization. This example showed that even a rudimentary visualization could make a microarchitectural performance problem visible enough to guide design changes.

\subsection{Catapult}
\begin{figure}[htbp]
    \centering
    \includegraphics[width=\columnwidth]{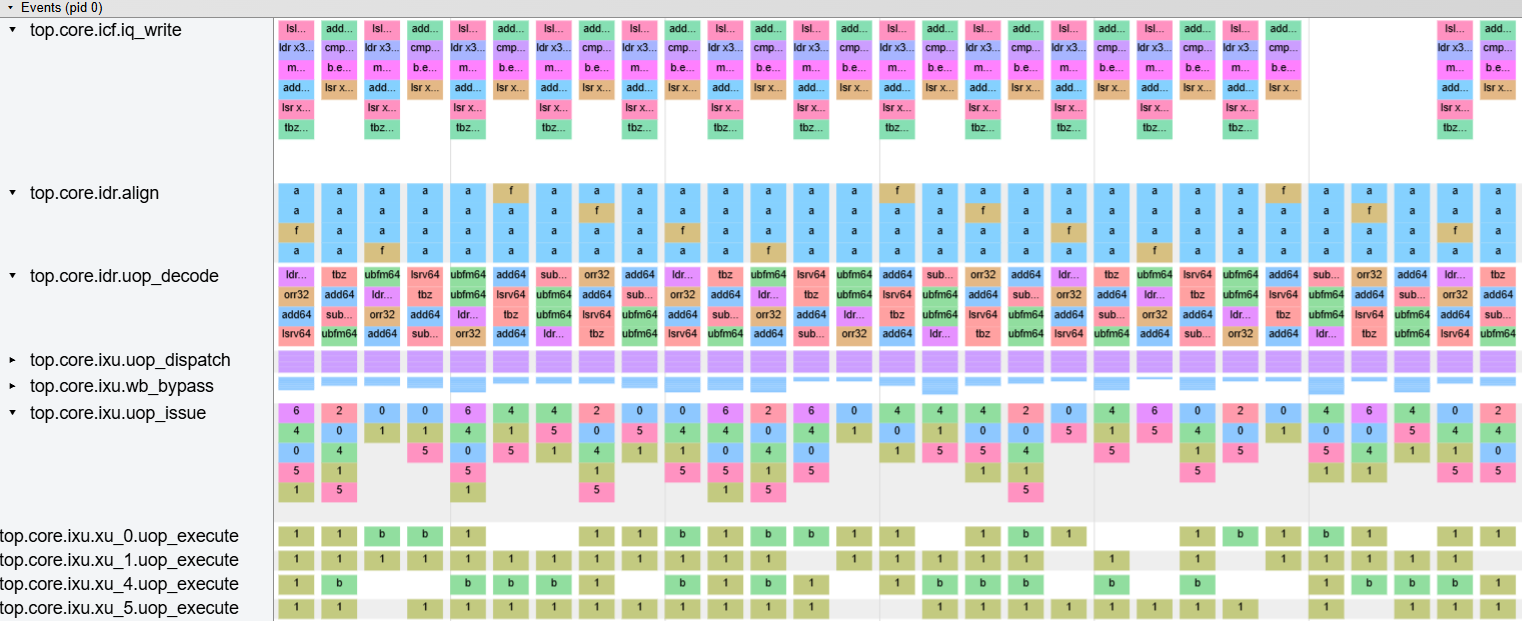}
    \caption{A snapshot of an event-stream visualized using the \textit{Chromium Catapult} interactive trace-viewer \cite{catapult_viewer}, with a resource-oriented view with cycles on the x-axis and various microarchitectural events on the y-axis. The sample above shows the pipeline running in a high IPC region of code, with the front-end delivering four uops per cycle to the scheduler and execution units.}
    \label{fig:catapult_pipeline}
\end{figure}

The initial text view was useful but severely limited. Navigation was slow, address and field search were not available, and there was no way to zoom out for a workload-level view. Rather than build a full custom viewer immediately, we adapted Catapult, a trace viewer originally built for browser profiling \cite{catapult_viewer}. We wrote a Python converter from event streams to Catapult's trace event format \cite{trace_event_format}. An example of a debug session using this technology is shown in \autoref{fig:catapult_pipeline}.

Catapult significantly improved interactivity with a relatively small investment of our team's effort, but its browser-profiling assumptions were limiting. Large numbers of rows were expensive, so the practical view was resource-oriented: each row represented an event type or resource rather than a single instruction or uop. Each event had an abbreviated name and associated values, and the values were visible when the user clicked on the event. Search was global and limited to exact string matching, however, so advanced or field-specific searches required workarounds.

Because it was a more natural fit for Catapult, resource view became the most common workflow. We also approximated a transaction view by wrapping transactions across a fixed number of rows; for example, instructions could be mapped to rows by ROB ID. This gave architects some of the benefits of a waterfall view, but overlapping transactions and configuration-dependent layout reduced its usability.

\begin{figure*}[ht!]
    \centering
    \includegraphics[width=\textwidth]{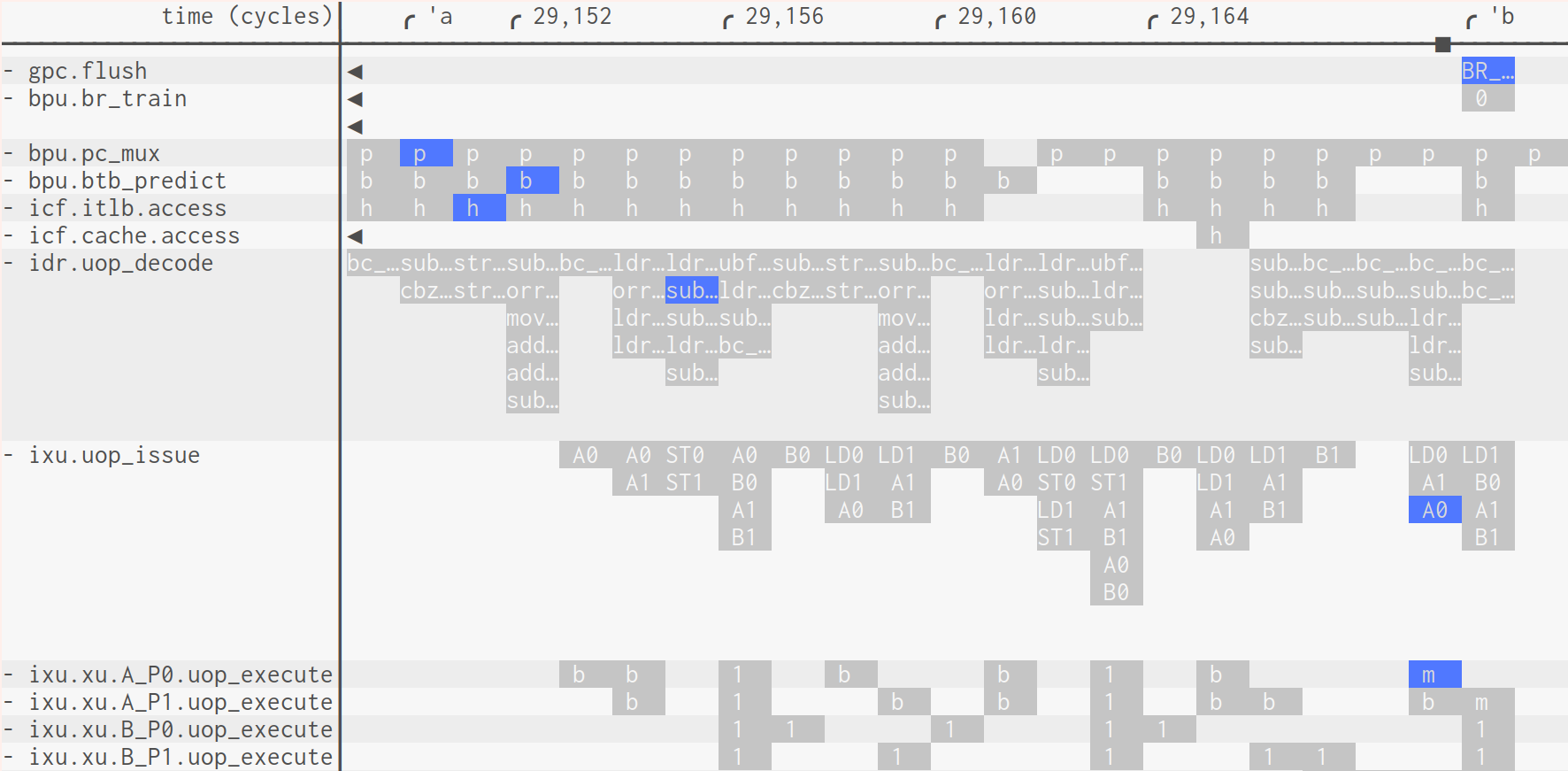}
    \cprotect\caption{Highlighting the transaction hierarchy in Catscan immediately links the flush (upper right, marked \verb|'b|), the execution of the mispredicted branch uop (lower right), and the original branch prediction (upper left, marked \verb|'a|). In this debug session, we discovered that this particular branch's PC had dropped some training events, which led to the misprediction; fixing the training logic led to an increase in prediction accuracy.}
    \label{fig:catscan_mispred_transaction_highlights}
\end{figure*}

\subsection{Visualization Philosophy}

Catapult also affected how the team reasoned about performance. Since its primary view was resource-centric (showing machine state and functional-unit occupancy over time) engineers who learned to debug with Catapult first naturally asked questions such as, ``Is the instruction queue full?'' or ``Are execution units starved?'' Other architects, who had used waterfall diagrams in earlier roles, tended to think transactionally, following one instruction and its dependencies through the pipeline. Both habits are useful, and the differences made it clear that a good viewer should support both styles of performance debug.

This split is not unique to our team. Intel had a similar divide between product architects working with the Keiko simulator and research architects working with ASIM \cite{asim_sim}. The ASIM environment used a resource-centric view called Dreams \cite{asim_dreams}, while the Keiko flow favored a transactional waterfall view through the jPTV pipetrace visualizer \cite{intel_jptv, arch_viz}. Later analysis showed that the two views were mathematically equivalent: jPTV data could be converted into resource views, and Dreams files could be transformed into waterfall graphics. The important lesson was that the views are complementary, not competing.

The less obvious lesson was about data representation and the inefficiency of conversion. The act of translating between two final, rigid formats led to the insight that the better approach would be for the simulator to output a flexible intermediate representation from which multiple views could be derived naturally. This principle was learned from the practical limitations of that era, and later motivated the Ampere event stream format. We designed it to be a single canonical source of pipeline behavior, flexible enough to power any visualization style or analysis tool. Using this event stream format from the beginning allowed us to easily explore and develop new analysis methods and visualization tools, ultimately fostering the development of Catscan.

\section{Catscan}
\label{sec:catscan}

\begin{figure*}
    \centering
    \includegraphics[width=\textwidth]{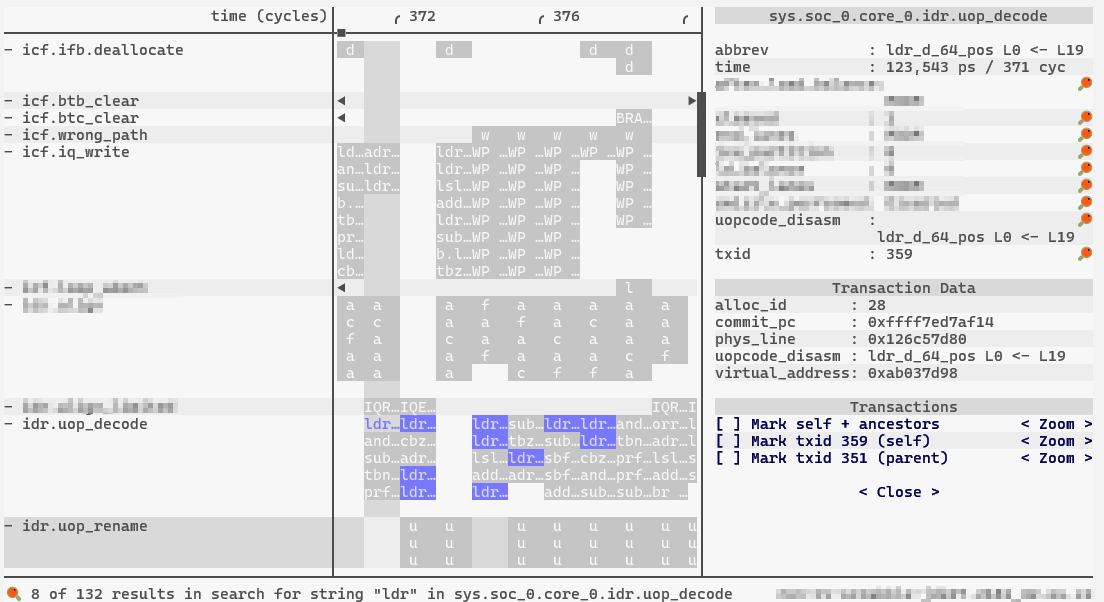}
    \cprotect\caption{Catscan search, showing matches for all uops containing the string ``ldr'' in the \verb|idr.uop_decode| row. In this performance debug session, we discovered that load uops on the wrong path were aliasing with correct path loads for L1 prefetcher training; the modeling was high-fidelity and unfortunately correct.}
    \label{fig:catscan_search}
\end{figure*}

\subsection{Motivation}
Catapult was a clear improvement over static text diagrams, but it was still a browser-profiling tool adapted to a CPU-debug problem. Its search was text-based, so architects used brittle workarounds such as injecting a synthetic string like \verb|txid_9403| into every event belonging to transaction 9403. It could not naturally answer domain-specific questions, such as finding all accesses to a cache line rather than exact address matches. It also had interaction problems for our workflow; for example, ordinary navigation dismissed search highlights, forcing users to recreate context repeatedly. Finally, Catapult was no longer an actively extensible base for our needs and was being superseded by Perfetto \cite{perfetto_viewer}. We needed a viewer designed around event streams and microarchitectural debug.

Catscan, a Python terminal user interface (TUI) for event streams, was written to answer this need. It is influenced by the efficiency of text editors such as Vim \cite{vim} and intended to move beyond the static text pipeline views common in academic simulators \cite{gem5_viz}. The terminal interface was a practical choice, as architects could run the tool directly on the compute servers where simulations completed. This avoided file transfers of large traces, while still getting interactive navigation with fast zooming, search, and selection. The name follows our internal convention of using cat-related names for performance model infrastructure, while reflecting the diagnostic nature of the tool.

\subsection{Features}

The initial goal for Catscan was to preserve the useful parts of the Catapult workflow while removing the limitations that slowed down performance debug. Owning the viewer also let us add new and more advanced features to further accelerate our work. The features that mattered most in production use fall into three groups: preserving causal context, keeping high-level performance signals visible while navigating detail, and supporting comparison between related traces. This section describes those features and the workflows they support.

\subsubsection{Transactional Exploration}

The most important early requirement was direct exploration of transaction relationships. Transactions are central to the event stream format, but in Catapult architects' exploration of them was limited to text search of injected metadata. After Catscan reached basic feature parity, transaction highlighting became one of the first domain-specific features.

When an event is selected, Catscan can highlight all events in the same transaction. The user can do this with Vim-style keyboard shortcuts or by selecting the relevant transaction in the event-detail sidebar with the mouse. Catscan can also highlight ancestor and descendant transactions, and multiple transactions or transaction groups can remain highlighted until they are cleared. For example, after selecting a branch uop that mispredicted at execution, the user can highlight the uop, its instruction, and the fetch transaction that produced it. This makes the original prediction, execution, and flush visible as one related chain, as shown in \autoref{fig:catscan_mispred_transaction_highlights}. When transactions are highlighted, unrelated events remain visible but are muted so the selected context stands out.

\subsubsection{Search}

Another early feature was advanced search. Catapult's text search was functional but too limited for microarchitectural exploration. Catscan added several capabilities that made common debug tasks faster:
\begin{enumerate}
    \item Masked integral search (i.e. search for integers matching \verb|0x45a83| for bits in mask \verb|0xfffc0|)
    \item Limiting search to specific events or fields by name or regex
    \item Searching based on min-occurrences in a cycle (useful for finding cases of peak throughput)
    \item Keeping search results highlighted until the user explicitly dismisses them
    \item Quickly searching for a field value within a focused event
\end{enumerate}

An example is shown in \autoref{fig:catscan_search}, where a user investigating a particular class of load uop at decode can type \verb|:search ldr rows=*.idr.uop_decode| to search for ``ldr'' while restricting the search to the \verb|*.idr.uop_decode| event. For address-oriented debugging, Catscan can also use an integer mask, e.g., \verb|:search 0x45a83 fields=*.address mask=0xfffc0|, which matches addresses with the selected cache-line bits.

\subsubsection{Summarization}

In resource view, events in a row may have different abbreviations (the short text shown inside the block representing the event). We use abbreviations to expose useful details without requiring the user to open every event. For example, a flush event might show whether it was caused by a branch misprediction or memory-disambiguation event, and a decode event might show the uop mnemonic. Catscan's \verb|:summarize| command reports a histogram of abbreviations or, with additional arguments, arbitrary event metadata.

\begin{figure}
    \centering
    \includegraphics[width=\columnwidth]{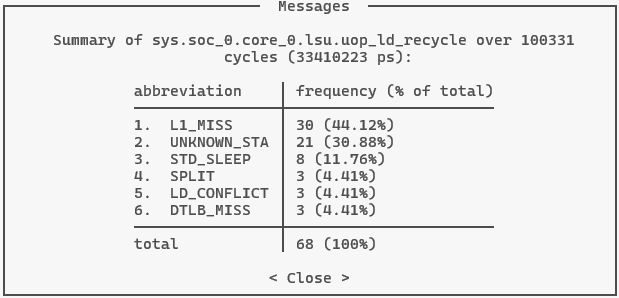}
    \caption{Catscan summarization of reasons for loads recycling. We encountered some outliers in performance verification with an uncharacteristically high number of UNKNOWN\_STA-based recycling, which led to the discovery of a logic signal not being hooked up. The store uop address was in fact known, but the communication was missing; fixing the bug alleviated the slowdown.}
    \label{fig:catscan_summarization}
\end{figure}

\autoref{fig:catscan_summarization} shows an example of typical use: marking the beginning and end of a burst of load recycling and running \verb|:summarize a b fields=*.reason| to ask whether the region is dominated by ordering replays, cache misses, or another recycle cause.

\subsubsection{Marking}

Microarchitectural debug often requires several events to be considered together. Without a way to mark important events and return to them, the user repeatedly searches for the same locations and loses context.

Catscan borrows Vim-style marks for this purpose. Marks are named with single characters which are set and revisited with short key sequences. Their locations are shown along the time bar at the top of the screen (see \verb|'a| and \verb|'b| in \autoref{fig:catscan_mispred_transaction_highlights}).

Marks also define time regions that other commands can operate on. For example, after setting marks at the start and end of a region with \verb|ma| and \verb|mb|, an architect can later use \verb|:zoom marks a b| to set the viewport to the time interval they define, use \verb|:summarize a b| to summarize events within that interval, or measure the time between them. This is representative of a broader Catscan design preference: a small number of composable features is more powerful than a larger set of isolated commands.

\subsubsection{Commit Synchronization}

Comparative analysis is a common performance-architecture task. Engineers often need to understand why a workload behaves differently across two models, two configurations, or two product generations. Synchronized dual-trace views have long existed in industrial tools, and academic work has also recognized their value for differential performance analysis \cite{dptv}. The need is especially acute during performance-verification correlation, where the goal is to explain differences between an RTL implementation and the performance model. Without integrated support, an architect typically opens two viewers and manually keeps them aligned to equivalent locations in the dynamic instruction streams.

Catscan can synchronize two event streams so the same portion of the workload is visible in each window, even if that portion occurs at different simulated times. The feature depends on unique markers emitted into both streams. In our flow, compatible models emit consistent IDs for committed instructions, and Catscan uses those IDs as waypoints. Moving either view moves the synchronized view accordingly, which removes a large amount of manual trace alignment during correlation work.

In addition to synchronizing the views, we use the combined data from both event streams to add ``virtual'' events to further aid architects in debugging. When debugging core-level performance, one of the top-level indicators of performance differences at the granularity of individual instructions is what we call ``commit push-out.'' An instruction's commit push-out is the number of cycles between when it commits and when the preceding instruction commits. We have found that when an architect knows a workload is behaving differently in two models or configurations, but not precisely where or why, studying the differences in commit push-out can help them quickly locate the cause. We insert a row of events which corresponds to the ``excess commit push-out'' relative to the other stream. When an instruction in the first event stream has a greater commit push-out than the same instruction in the second stream (identified by a consistent ID in the two traces), an event is inserted for each cycle of the difference. When the same instruction has the same commit push-out across both traces, no virtual event is inserted. These virtual events are inserted with the transaction ID of the relevant instruction to allow the virtual event to be correlated with the relevant instruction. These virtual events help identify divergent instruction behavior (for example, due to cache misses in one stream which do not occur in the other).

\begin{figure}
    \centering
    \includegraphics[width=\columnwidth]{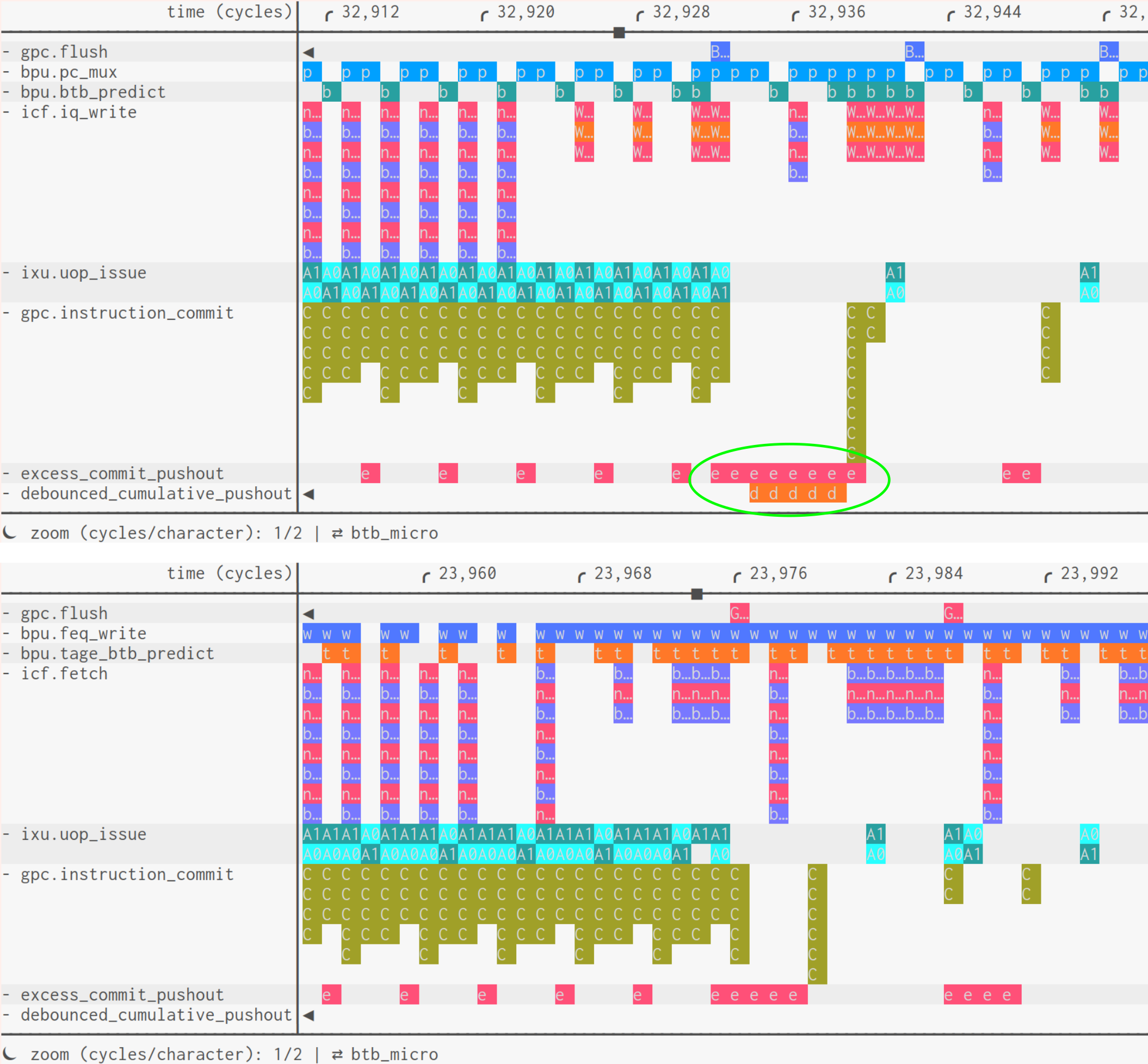}
    \cprotect\caption{An example of using Catscan's commit-syncing functionality to debug event streams of the same workload on our software performance model (above) and in RTL simulation (below). The virtual \verb|debounced_cumulative_pushout| events (|\allochighlightnarrow{d}| in green oval) call attention to a location where the sum of the excess commit push-out (|\misshighlightnarrow{e}|) diverges beyond the allowed window---a potential spot for further human debug.}
    \label{fig:catscan_commit_sync}
\end{figure}

Straightforward excess commit push-out is not always enough. In some workloads, small microarchitectural differences cause excess push-out to alternate rapidly between the two streams. The result is many scattered virtual events that obscure the larger divergence. We tried several derived signals, including moving averages, before settling on ``de-bounced cumulative push-out.'' The algorithm allows a sliding window in the two streams' push-out difference and inserts virtual events only when the window must move to keep bounding the current difference. This suppresses transient differences while highlighting regions where performance has begun to diverge materially. \autoref{fig:catscan_commit_sync} shows a pair of streams where excess commit push-out produces many spurious events, while de-bounced cumulative push-out identifies a smaller region for further debug.

More generally, we have found it useful to compute virtual events from both sides of a comparison and display those events as if they had come from the original streams.

\begin{figure*}
    \centering
    \includegraphics[width=\textwidth]{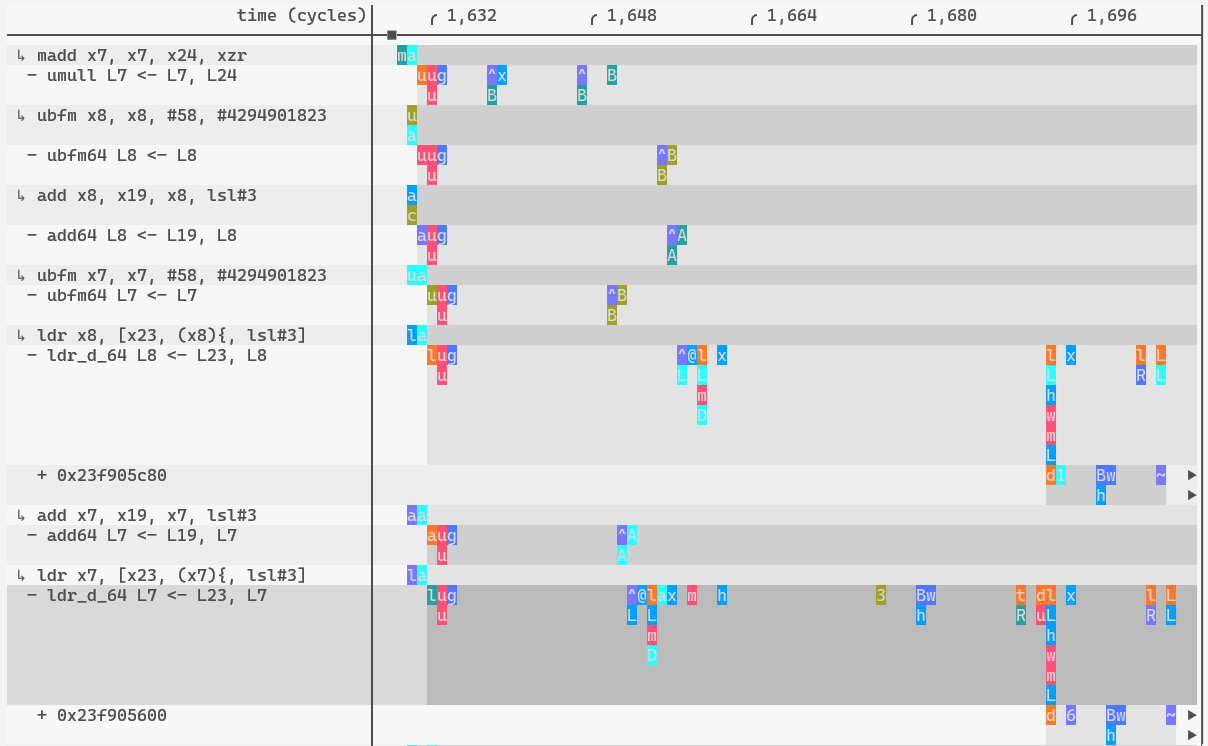}
    \caption{Catscan with transaction-oriented view showing instructions, uops, and cache misses. We see two load uops that each miss the L1 cache |\misshighlightnarrow{m}|, spawn transactions by allocating fill buffers for demand requests |\allochighlightnarrow{d}|, which hit in the L2 six cycles later |\hithighlightnarrow{h}| and eventually return data to the L1 and deallocate |\deallochighlightnarrow{\textasciitilde}|.}
    \label{fig:catscan_transaction_view}
\end{figure*}

\subsubsection{Row Pinning}

In resource view, Catscan displays one row for each requested event type and lets the user scroll through those rows. This flexibility is important because the path of a debug session is rarely known in advance. Over time, however, our default configurations grew to include many rows. Keeping many events available avoided stopping a debug session to collect or load more data, but it also made important context easy to lose off screen.

To address this, Catscan can pin selected rows to the top of the window so they remain visible while other rows scroll. Any row can be pinned, but the feature is most useful for high-level signals such as CPU stalls, as shown in \autoref{fig:catscan_pinning}. Pinned rows are displayed in condensed form: they omit event abbreviations and occupy one terminal row, preserving vertical space while keeping overview signals visible.

\subsubsection{Transaction View}

Although resource view remains our default, Catscan also provides transaction-oriented views. Time still progresses left to right, but each row contains the events within one transaction rather than all events of one type. This organization makes transaction latency easier to inspect and helps expose why an instruction, uop, or memory operation was delayed.

Event streams' transaction graph also improves this view. Catscan can nest descendant transactions under their parent, shown by row indentation in \autoref{fig:catscan_transaction_view}. This makes spawned work visible; for example, a load uop can be shown with the memory transaction created after an L1 miss. In practice, the two views serve different purposes. Resource view gives a quick system-level overview, while transaction view exposes instruction- and uop-level delay patterns.

\subsubsection{Movement and UI}

Because Catscan is terminal-based and influenced by Vim, keyboard navigation is the fastest path for experienced users. Examples of common motions include \verb|hjkl| for moving through events and rows and \verb|Ctrl-u| for moving up half a page. We also added mouse interaction to reduce the adoption cost for new users. Panning, zooming, and selection follow common browser and image-editor conventions, while event-viewer-specific interactions include double-clicking to jump to the nearest event in a sparse row.

Catscan also includes a Vim-style command prompt with history and tab completion. The prompt provides a place for commands that are more global than a local key binding, such as:
\begin{itemize}
    \item \verb|:2400| (cycles) or \verb|:7200ps| jump to known times, for when an ``interesting'' time is known.
    \item \verb|:*.instruction_commit| jumps to the next matching event row from a glob, supporting quick navigation.
    \item \verb|:zoom search|, \verb|:zoom highlights|, and \verb|:zoom marks a b| resize and center the view around the requested context and help show its relevant ``lifetime.''
\end{itemize}

\subsection{Internal Design}

In addition to the user-visible features, several internal design choices were important to Catscan's responsiveness and may apply to other microarchitectural visualization tools.

\subsubsection{Data Structures}

\begin{figure}
\centering
\begin{tikzpicture}[
  font=\scriptsize,
  node distance=4mm,
  box/.style={
    draw, rounded corners, align=center,
    minimum width=6.6cm, inner xsep=4pt, inner ysep=3pt
  },
  halfbox/.style={
    draw, align=left, minimum width=3.0cm, inner xsep=4pt, inner ysep=3pt, font=\scriptsize
  },
  arr/.style={-{Latex[length=1mm,width=1mm]}, line width=0.6pt},
  dashedarr/.style={dotted, -{Latex[length=1mm,width=1mm]}, line width=0.6pt},
  note/.style={font=\scriptsize, align=center}
]

\node[box] (esd) {
  \texttt{\textbf{EventStreamData}} \\ \texttt{dict[name: EventData]}
};

\node[box, below=6mm of esd] (ed) {
  \texttt{\textbf{EventData}} \\ \texttt{times: array (sorted)} \\ \texttt{events\_by\_time: dict[time: list[Event]]}
};

\coordinate (edSW) at (ed.south west);
\coordinate (edSE) at (ed.south east);

\node[halfbox, anchor=north west] (times) at ($(edSW)+(0,-8mm)$) {
  \textbf{\texttt{times} (sorted)}:\\ \texttt{[ t0, t1, t2, \ldots, tN ]}
};

\node[halfbox, anchor=north east] (dictt) at ($(edSE)+(0,-8mm)$) {
  \textbf{\texttt{events\_by\_time} (dict)}:\\ \texttt{t0 : [Event0, Event1]} \\ \texttt{t1 : [Event2]} \\ \texttt{t2 : [Event3, Event4]} \\ \texttt{\ldots}
};

\draw[arr] (ed.south) -- (times.north);
\draw[arr] (ed.south) -- (dictt.north) node[midway, right=10pt, font=\scriptsize] {same times as keys};

\draw[arr] (esd.south) -- (ed.north) node[midway, right=2pt, font=\scriptsize] {rows indexed by name};

\node[note, below=4mm of times] (bs) {\emph{Binary search on the sorted times}\\\emph{array for efficient locate / range }\\\emph{iterate}};
\draw[dashedarr] (bs.north) -- (times.south);

\end{tikzpicture}
\caption{Catscan event data structure organization}
\label{fig:catscan_data_structures}
\end{figure}
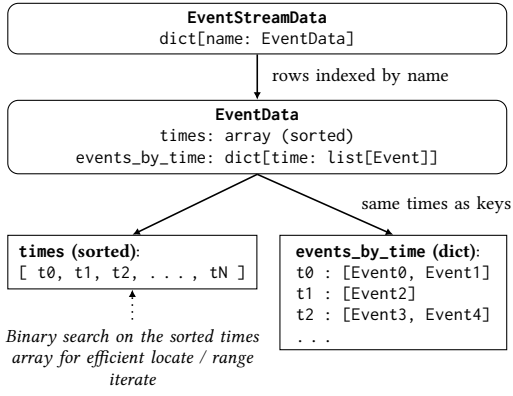

The first choice was to align data structures with the queries we expected to be common and expensive. We observed high temporal locality in Catscan's queries of time ranges and event rows, because the user normally focuses on a relatively small portion of the stream at a time. To take advantage, at the root \verb|EventStreamData| stores rows of events (\verb|EventData| instances) in a Python dictionary indexed by row name. Each \verb|EventData| instance stores the times at which events occur in a sorted Python array, plus a dictionary from timestamps to the list of events at that time. This structure, shown in \autoref{fig:catscan_data_structures}, supports binary search to locate a time and efficient iteration over a time range.

\subsubsection{Dithering}

At high zoom levels Catscan can display individual events, but when zoomed out it must represent many events with one terminal character. Catscan uses Unicode block element characters \cite{blockelements} and varies the displayed height according to the number of events represented by that character. However, if each character is rendered independently, information can be lost because only a fixed number of heights and colors are available. Two regions with different average event densities may map to the same displayed height.

This resembles the graphics problem of displaying an image using fewer colors than the original. A common solution is error diffusion, including the Floyd-Steinberg dithering algorithm \cite{floyd_steinberg}. Error diffusion adjusts local displayed values so the larger region better represents the source, even if an individual pixel or character is farther from its ``true'' value. That tradeoff matches our use case: when zoomed out, an architect is usually looking for larger patterns rather than the exactly-correct value of one time granule.

\begin{figure}
    \centering
    \includegraphics[width=\columnwidth]{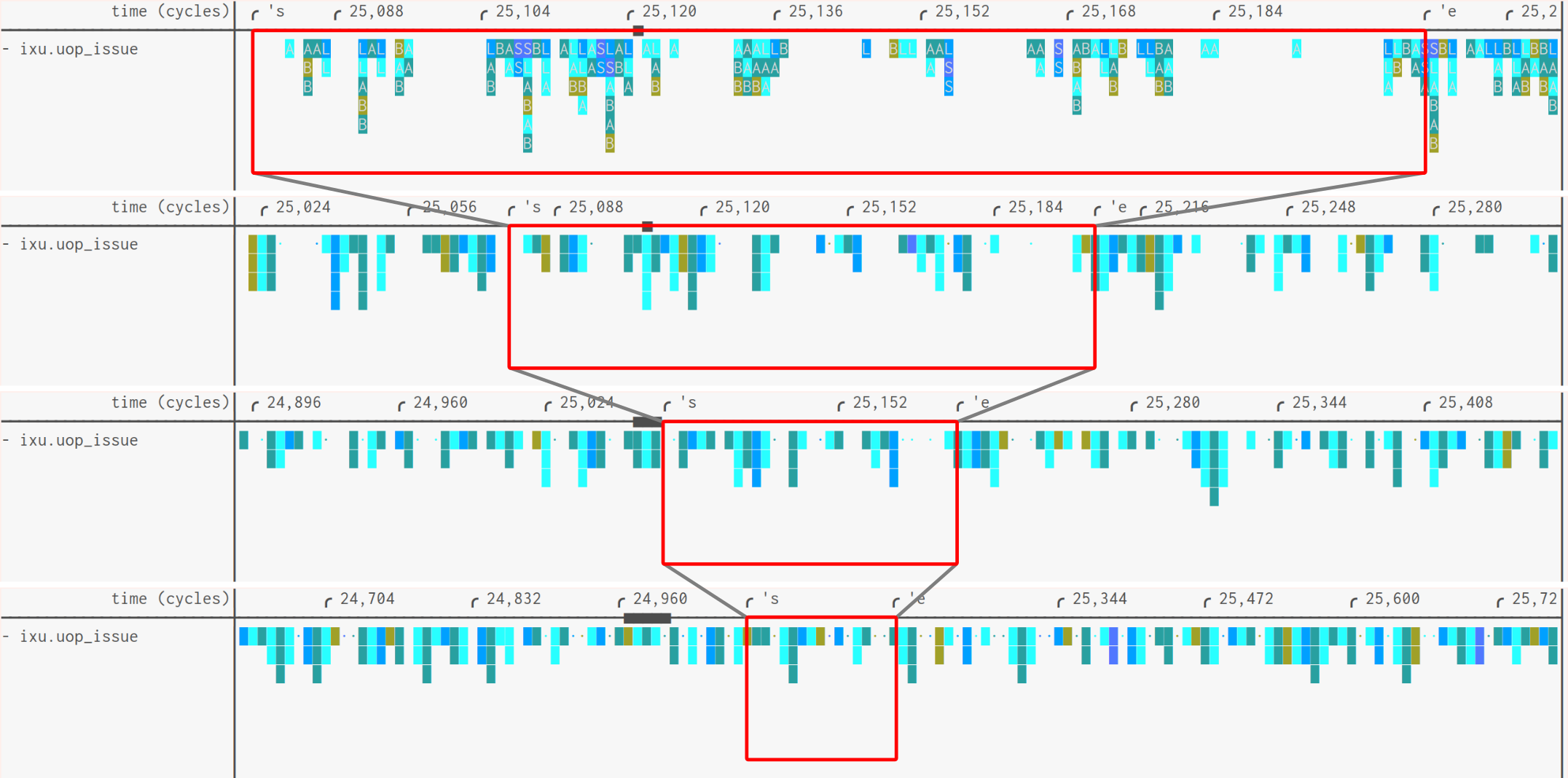}
    \caption{Zooming out on the same region of a row of events (highlighted with red boxes) shows how Catscan's error diffusion continues to represent the frequency and colors of the underlying events within the available resolution.}
    \label{fig:catscan_dithering}
\end{figure}

Catscan therefore applies error diffusion while rendering event rows. As a row is processed from left to right, the algorithm tracks residual error for character height and for each color. At each character position, Catscan counts the events mapped to that character at the current zoom level and counts how many of them have each color. Residual error is added to those counts, then the closest representable height and color are chosen. The displayed height and color are subtracted from the adjusted counts, and the residual is carried forward. \autoref{fig:catscan_dithering} shows the same region rendered at several zoom levels with this strategy.

We added two small adjustments for usability. First, residual color error could cross an empty region and cause a color to appear on the far side even though that color was absent there. Catscan applies an error-falloff factor whenever a character represents no events, reducing that artifact. Second, sparse regions could render as zero-height bars even though events were present. When a character represents a nonzero number of events but would otherwise render as zero height, Catscan draws a period. This tells the user that something occurred at that location while still indicating that the density is below the lowest bar height.

\subsubsection{Caching}

Catscan uses caching and partitioning to keep interaction responsive. A zoomed-out view may require processing hundreds of thousands of events, and smooth panning requires rendering new frames many times per second. Since event streams are static once loaded, caching is natural, but caching complete screens is ineffective: zoom level, time window, displayed rows, terminal size, highlights, and other state create too many combinations.

Instead, Catscan caches intermediate rendering results: character heights and colors for sections of event rows. It also partitions event time into power-of-two-aligned regions, similar in spirit to spatial pre-partitioning used by rendering engines like Doom \cite{doom}. When rendering a viewport, Catscan requests the relevant pre-rendered regions for each row, reuses cached regions when available, and cuts and splices them to match the current view. This combination of partitioning and intermediate-result caching lets Catscan render dense event streams while preserving interactive navigation.

\subsection{Limitations}

Catscan currently reads and preprocesses an event stream in its entirety before visualization. Event streams are encoded as variable-length records and currently lack a random-access index, so Catscan cannot jump directly to an arbitrary time range without first reading the stream. This creates startup and memory costs that bound the practical size of traces loaded interactively.

In our workflow, this has been acceptable because Catscan is usually not the first pass over a long simulation. Architects use aggregate metrics, summaries, and filters to identify a region of interest, then load a detailed event stream limited to that smaller region. We typically inspect windows on the order of tens of thousands of instructions, often around 50,000 instructions, rather than full benchmark traces. If larger interactive traces become important, the most direct extension would be to add an index mapping time ranges to positions in the event stream so Catscan can load and preprocess selected regions on demand.

Being a TUI also constrains the visualization. Rendering is limited to terminal cells, colors, and Unicode block characters, and the interaction model is less rich than a desktop or web viewer. Catscan mitigates this with compact row rendering, dithering, persistent highlights, keyboard commands, and limited mouse support, but the TUI remains a tradeoff.

\section{Conclusion}

We have presented event streams and Catscan, two pieces of performance-analysis infrastructure developed for Ampere's CPU design flow. Event streams provide a compact, flexible, and implementation-neutral way to record typed microarchitectural events together with transaction relationships. Software models, RTL simulations, and emulation can emit the same format, letting architects use a common analysis stack across sources. Catscan uses this format to support both resource- and transaction-oriented views, provide useful context, domain-specific search, summarization, and synchronized trace comparison. Together, these tools help efficiently identify performance symptoms and the specific instruction, uop, resource, or transaction chain that explains them.

Our experience demonstrates it is useful for the trace format to be treated as a canonical intermediate representation rather than as input to one viewer or tool. The event stream format we describe has proved flexible enough to derive complementary views, power a variety of analysis tools, and preserve causal context across debug workflows. The main remaining scaling tradeoff points to a natural future extension: adding indexing would allow Catscan to provide usable interactive visualization beyond the focused regions of interest we inspect today. By releasing both the event-stream format and Catscan, we hope to give academic and industrial users the means to adapt these workflows to their own simulators and design environments.

\bstctlcite{IEEEtran_bst_ctl}

\bibliographystyle{IEEEtran}
\bibliography{90-citations}

@IEEEtranBSTCTL{IEEEtran_bst_ctl,
  CTLuse_url = {yes},
  CTLdash_repeated_names = {no}
}

@inproceedings{ampere_pv,
  author       = {Doa'a Al-Otoom and Nicholas Kelly and others},
  title        = {{Performance Verification of the AmpereOne® CPU Core}},
  booktitle    = {International Symposium on Microarchitecture (MICRO-59), Athens, Greece},
  publisher    = {{IEEE}},
  year         = {2026},
  eprint={2608.19300},
  archivePrefix={arXiv},
  primaryClass={cs.AR},
  url={https://arxiv.org/abs/2608.19300}, 
}

@INPROCEEDINGS{ampere_mte,
      title={{Optimized Memory Tagging on AmpereOne® Processors}},
      author={Shiv Kaushik and Mahesh Madhav and others},
      year={2025},
      eprint={2511.17773},
      archivePrefix={arXiv},
      primaryClass={cs.AR}, 
      publisher = {IEEE},
      booktitle = {Proceedings of the 53rd Annual International Symposium on Computer Architecture},
      series = {ISCA '26},
      location = {Raleigh, NC, USA},
      doi = {10.1109/ISCA66397.2026.00080},
      url={https://doi.org/10.1109/ISCA66397.2026.00080}
}

@inproceedings{ampere_panem,
      title={{PANEM: A Heuristic Latency Model}}, 
      author={Lucas Crowthers and Mahesh Madhav},
      year={2026},
      eprint={2609.29629},
      archivePrefix={arXiv},
      primaryClass={cs.AR},
      booktitle = {2026 IEEE International Symposium on Workload Characterization},
      series = {IISWC '26},
      location = {Boulder, CO},
      url={https://arxiv.org/abs/2609.29629}, 
}

@ARTICLE{alpha_sim,
    author={Edmondson, John and Reilly, Matt},
    journal={Computer},
    title={{Performance Simulation of an Alpha Microprocessor}},
    year={1998},
    volume={31},
    number={05},
    ISSN={1558-0814},
    pages={50-58},
    doi={10.1109/2.675634},
    url={https://doi.org/10.1109/2.675634},
    publisher={IEEE Computer Society},
    address={Los Alamitos, CA, USA},
    month=may
}

@ARTICLE{floyd_steinberg,
    author={Floyd, R.W. and Steinberg, L.},
    journal={Proceedings of the Society of Information Display},
    title={{An Adaptive Algorithm for Spatial Grey Scale}},
    year={1976},

    url={https://en.wikipedia.org/wiki/Floyd-Steinberg_dithering}
}

@misc{asim_dreams,
    author = {Pau Cabre and Mohit Gambhir and Joel Emer},
    year = {2015},
    title = {{ASIM-Dreams/dreams}},
    url = {https://github.com/Asim-Dreams/dreams}
}

@misc{catscan_repo,
    author = {{Ampere Computing}},
    year = {2026},
    title = {{Catscan Source Repository}},
    url = {https://github.com/AmpereComputing/catscan}
}

@misc{perf_streams_repo,
    author = {{Ampere Computing}},
    year = {2026},
    title = {{Performance Streams Source Repository}},
    url = {https://github.com/AmpereComputing/perf-streams}
}

@misc{protobuf,
    author = {Google},
    year = {2026},
    title = {{Protocol Buffers}},
    url = {www.protobuf.dev}
}

@misc{catapult_viewer,
    author = {Google},
    year = {2026},
    title = {{Catapult Trace Viewer}},
    url = {https://chromium.googlesource.com/catapult/+/HEAD/tracing/README.md}
}

@misc{perfetto_viewer,
    author = {Google},
    year = {2026},
    title = {{Perfetto Trace Viewer}},
    url = {https://perfetto.dev/docs/}
}

@phdthesis{illinois_tracevis,
  title={{TraceVis: an Execution Trace Visualization Tool}},
  author={Roberts, James Evan and Zilles, Craig},
  year={2004},
  school={University of Illinois},
  url={https://zilles.cs.illinois.edu/papers/tracevis.msthesis.pdf}
}

@INPROCEEDINGS{michigan_pipeviz,
  author={Chris Weaver and others},
  booktitle={2001 IEEE International Symposium on Performance Analysis of Systems and Software. ISPASS.}, 
  title={{Performance Analysis Using Pipeline Visualization}}, 
  year={2001},
  volume={},
  number={},
  pages={18-21},
  url = {https://doi.org/10.1109/ISPASS.2001.990670},
  doi={10.1109/ISPASS.2001.990670}
}

@inproceedings{stanford_rivet,
  author       = {Chris Stolte and
                  Robert Bosch and
                  Pat Hanrahan and
                  Mendel Rosenblum},
  title        = {{Visualizing Application Behavior on Superscalar Processors}},
  booktitle    = {{IEEE} Symposium on Information Visualization},
  address      = {San Francisco, CA},
  series       = {INFOVIS '99},
  pages        = {10--17},
  publisher    = {{IEEE} Computer Society},
  year         = {1999},
  url          = {https://graphics.stanford.edu/papers/rivet_pipeline/},
  doi          = {10.1109/INFVIS.1999.801852},
}

@article{stanford_rivet2,
    author = {Bosch, Robert and Stolte, Chris and Tang, Diane and Gerth, John and Rosenblum, Mendel and Hanrahan, Pat},
    title = {{Rivet: a Flexible Environment for Computer Systems Visualization}},
    year = {2000},
    issue_date = {February 2000},
    publisher = {Association for Computing Machinery},
    address = {New York, NY, USA},
    volume = {34},
    number = {1},
    issn = {0097-8930},
    doi = {10.1145/563788.604455},
    url = {https://doi.org/10.1145/563788.604455},
    journal = {SIGGRAPH Comput. Graph.},
    month = feb,
    pages = {68--73},
    numpages = {6}
}

@phdthesis{stanford_bosch,
  title={{Using Visualization to Understand the Behavior of Computer Systems}},
  author={Bosch, Robert Peter and Rosenblum, Mendel},
  year={2001},
  school={Stanford University},
  url={https://dl.acm.org/doi/10.5555/934519},
}

@ARTICLE{asim_sim,
  author={Emer, J. and others},
  journal={Computer}, 
  title={{ASIM: a Performance Model Framework}}, 
  year={2002},
  volume={35},
  number={2},
  pages={68-76},
  url={https://doi.org/10.1109/2.982918},
  doi={10.1109/2.982918}
}

@online{arch_viz,
  author = {Rustam Miftakhutdinov},
  title = {{Memory-centric microarchitecture visualization}},
  year = {2013},
  month = aug,
  day = {14},
  url = {https://comparch.net/2013/08/14/memory-centric-microarchitecture-visualization/},
  urldate = {2026-05-07}
}

@online{gem5_viz,
  author = {{gem5 community}},
  title = {{gem5 O3 Pipeline Viewer}},
  url = {https://www.gem5.org/documentation/general_docs/cpu_models/visualization}
}

@online{vim,
  author = {Bram Moolenaar},
  title = {{Vim - the ubiquitous text editor}},
  url = {https://www.vim.org}
}

@inproceedings{dptv,
    author = {Adam Grunwald and Phuong Nguyen and Elliott Forbes},
    title = {{dptv: A New PipeTrace Viewer for Microarchitectural Analysis}},
    year = {2023},
    address = {Cedar Falls, IA, USA},
    booktitle = {Proceedings of the 55th Midwest Instruction and Computing Symposium},
    pages = {274-284},
    url={https://micsymposium.org/mics2023/wp-content/uploads/2023/04/Proceeding_MICS_2023_v2.pdf},
    url2={https://cs.uwlax.edu/~eforbes/dptv/},
    series = {MICS '23}
}

@inproceedings{satsim,
    author = {Wolff, Mark and Wills, Linda},
    title = {{SATSim: A Superscalar Architecture Trace Simulator Using Interactive Animation}},
    year = {2000},
    isbn = {9781450347303},
    publisher = {Association for Computing Machinery},
    address = {New York, NY, USA},
    doi = {10.1145/1275240.1275249},
    url = {https://doi.org/10.1145/1275240.1275249},
    booktitle = {Proceedings of the 2000 Workshop on Computer Architecture Education},
    pages = {6-es},
    series = {WCAE '00}
}

@online{doom,
  author = {Target, Sinclair},
  title = {{How Much of a Genius-Level Move Was Using Binary Space Partitioning in Doom?}},
  year = {2019},
  month = nov,
  day = {6},
  url = {https://twobithistory.org/2019/11/06/doom-bsp.html},
  urldate = {2026-05-05}
}

@online{jviewer,
  author = {Clarke, Paul},
  title = {{Porting and tuning tools for Linux on Power}},
  year = {2021},
  month = nov,
  day = {10},
  url = {https://developer.ibm.com/tutorials/l-porting-tuning-tools/#jviewer},
  urldate = {2026-05-19}
}

@online{intel_jptv,
  author = {Doug Carmean},
  title = {{The Intel® Pentium® 4 Processor}},
  note = {Presentation on Performance Analysis to Cornell University},
  year = {2005},
  month = apr,
  day = {26},
  url = {http://cwcyrix.nsupdate.info/ftp-archives/download.intel.com/education/highered/research/carmean_slides.pdf},
  urldate = {2026-05-19}
}

@misc{blockelements,
  author    = {{The Unicode Consortium}},
  title     = {{Block Elements}},
  howpublished = {The Unicode Standard, Version 17.0.0},
  year      = {2025},
  url       = {https://www.unicode.org/charts/PDF/U2580.pdf},
  urldate   = {2026-05-20},
}

@misc{trace_event_format,
  author    = {{Google; nduca@, dsinclair@}},
  title     = {{Trace Event Format}},
  year      = {2016},
  url       = {https://docs.google.com/document/d/1CvAClvFfyA5R-PhYUmn5OOQtYMH4h6I0nSsKchNAySU/preview},
  urldate   = {2026-05-21},
}


\end{document}